\documentclass[preprintnumbers,amsmath,amssymb,floatfix,11pt,prd,onecolumn,showpacs,
superscriptaddress,nofootinbib]{revtex4}
\usepackage{graphicx}
\usepackage{epsfig}
\usepackage{bm}
\usepackage{amsfonts}

\begin{document}

\title{\large{\bf HOW EFFECTIVE IS NEW VARIABLE MODIFIED CHAPLYGIN GAS TO PLAY THE ROLE OF DARK ENERGY-
A DYNAMICAL SYSTEM ANALYSIS IN RS II BRANE MODEL}}

\author{\textbf{Prabir Rudra}}\email{prudra.math@gmail.com}
\affiliation{Department of Mathematics, Pailan College of
Management and Technology, Bengal Pailan Park, Kolkata-700 104,
India.}
\author{\textbf{Chayan Ranjit}}\email{chayanranjit@gmail.com} \affiliation{Department of Mathematics, Seacom Engineering College, Howrah-711 302, India}
\author{\textbf{Sujata Kundu}}\email{sujatakundu10@gmail.com} \affiliation{Department of Information Technology, Narula Institute of Technology, Kolkata-700109,India}

\begin{abstract}
Motivated by some previous works of Rudra et al we set to explore
the background dynamics when dark energy in the form of New
Variable Modified Chaplygin gas is coupled to dark matter with a
suitable interaction in the universe described by brane cosmology.
The main idea is to find out the efficiency of New variable
modified Chaplygin gas to play the role of DE. As a result we
resort to the technique of comparison with standard dark energy
models. Here the RSII brane model have been considered as the
gravity theory. An interacting model is considered in order to
search for a possible solution of the cosmic coincidence problem.
A dynamical system analysis is performed because of the high
complexity of the system . The statefinder parameters are also
calculated to classify the dark energy model. Graphs and phase
diagrams are drawn to study the variations of these parameters and
get an insight into the effectiveness of the dark energy model. It
is also seen that the background dynamics of New Variable Modified
Chaplygin gas is consistent with the late cosmic acceleration.
After performing an extensive mathematical analysis, we are able
to constrain the parameters of new variable modified Chaplygin gas
as $m<n$ to produce the best possible results. Future
singularities are studied and it is found that the model has a
tendency to result in such singularities unlike the case of
generalized cosmic Chaplygin gas. Our investigation leads us to
the fact that New Variable Modified Chaplygin gas is not as
effective as other Chaplygin gas models to play the role of dark
energy.

\end{abstract}

\maketitle

\newpage

\section{INTRODUCTION}

\noindent

Recent cosmic acceleration is a well-known and accepted fact in
the cosmological society currently\cite{Perlmutter1, Spergel1}.
The root cause for this phenomenon is still under research.
However of late there have been some speculations regarding the
existence of a mysterious negative pressure component which
violates the strong energy condition i.e. $\rho+3p<0$. Because of
its invisible nature this energy component is aptly termed as dark
energy (DE) \cite{Riess1}.

\noindent

Since the concept of DE flourished in the last decade,
cosmologists all over the world started searching for a suitable
model of DE. As a result various DE models have come into
existence of late. DE represented by a scalar field \footnote{ in
the presence of a scalar field the transition from a universe
filled with matter to an exponentially expanding universe is
justified } \cite{Nojiri1} is often called quintessence. Not only
scalar field but also there are other Dark fluid models like
Chaplygin gas which plays the role of DE very efficiently. As time
passed extensive research was conducted and Chaplygin gas (CG)
\cite{Kamenshchik1, Gorini1}, got modified into Generalized
Chaplygin gas (GCG) \cite{Gorini2, Alam1, Bento1, Barreiro1,
Carturan1} and then to Modified Chaplygin gas (MCG)
\cite{Benaoum1, Debnath1}. In this context it is worth mentioning
that dynamics of MCG in Braneworld was studied by Rudra et al
\cite{Rudra1}. Other than these other forms of Chaplygin gas
models have also been proposed such as Variable Modified Chaplygin
gas (VMCG) \cite{Debnath2}, New Variable Modified Chaplygin gas
(NVMCG) \cite{Chakraborty1}, generalized cosmic Chaplygin gas
(GCCG) \cite{Gonzalez1}. Dynamics of GCCG in Loop Quantum
cosmology was studied by Chowdhury et al in \cite{Chowdhury1}. The
dynamics of GCCG in braneworld was studied by Rudra in
\cite{Rudra2}. Other existing forms of DE are phantom
\cite{Nojiri2}, k-essence \cite{Bamba1}, tachyonic field
\cite{Nojiri3}, etc.

\noindent

The equation of state for NVMCG is given by,
\begin{equation}
p=A(a)\rho-\frac{B(a)}{\rho^{\alpha}}~,~~~~~~~~~~0\leq \alpha \leq
1
\end{equation}
Here we will consider $A(a)=A_{0}a^{-n}$ and $B(a)=B_{0}a^{-m}$,
where $A_{0}$, $B_{0}$, $\alpha$, $m$ and $n$ are positive
constants.

\noindent

Currently, we live in a special epoch where the densities of DE
and DM are comparable. The fact that they have evolved
independently from different mass scales makes the fact more
interesting. Given their non co-existence in evolution, comparable
densities is quite an unexpected phenomenon. This is known as the
famous cosmic coincidence problem. Till date several attempts have
been made to find a solution to this problem \cite{del Campo1,
Berger1}. A suitable interaction between DE and DM provides the
best method of solution for this problem. It is obvious that a
transition has occurred from a matter dominated universe to dark
energy dominated universe, by exchange of energy at an appropriate
rate. Now in order to be consistent with the expansion history of
the universe as confirmed by the supernovae and CMB data
\cite{Jamil1} the decay rate has to be fixed such that it is
proportional to the present day Hubble parameter. Keeping the fact
in mind cosmologists all over the world have studied and proposed
a variety of interacting DE models \cite{Setare1, Setare2, Hu1,
Wu1, Jamil2, Dalal1}.

\noindent

Now, dark energy is not the only concept that can demonstrate the
present day universe. The left hand side of the Einstein's field
equation can also modified, to obtain suitable results. This
modification however gives rise to the famous modified gravity
theories, which in their own right can independently give us
suitable models for our expanding universe. In this context
Brane-gravity was introduced and brane cosmology was developed. A
review on brane-gravity and its various applications with special
attention to cosmology is available in \cite{Maartens1, Rubakov1,
Brax1}. In this work we consider a very popular model of brane
gravity, namely the RS II brane. The main objective of this work
is to examine the nature of the different physical parameters of
the DE for the universe around the stable critical points in the
brane model in presence of NVMCG. Effectiveness and success of the
mathematical formulation of NVMCG will be studied. Impact of any
future singularity caused by the DE in the brane world model will
also be studied.

\noindent

This paper is organized as follows: Section 2 comprises of the
analysis in RS II brane model. In section 3, a detailed graphical
analysis for the phase plane is given. In section 4, some details
regarding the mathematical construction of NVMCG is provided. In
section 5 future singularities arising from the model are studied
and finally the paper ends with some concluding remarks in section
6.

\section{MODEL 1: RS II BRANE MODEL}

\noindent

Randall and Sundrum \cite{Randall1, Randall2} proposed a
bulk-brane model to explain the higher dimensional theory,
popularly known as RS II brane model. According to this model we
live in a four dimensional world (called 3-brane, a domain wall)
which is embedded in a 5D space time (bulk). All matter fields are
confined in the brane whereas gravity can only propagate in the
bulk. The consistency of this brane model with the expanding
universe has given popularity to this model of late in the field
of cosmology.

\noindent

In RS II model the effective equations of motion on the 3-brane
embedded in 5D bulk having $Z_{2}$-symmetry are given by
\cite{Maartens1, Randall2, Shiromizu1, Sasaki1, Maeda1}
\begin{equation}
^{(4)}G_{\mu\nu}=-\Lambda_{4}q_{\mu\nu}+\kappa^{2}_{4}\tau_{\mu\nu}+\kappa^{4}_{5}\Pi_{\mu\nu}-E_{\mu\nu}
\end{equation}
where

\begin{equation}
\kappa^{2}_{4}=\frac{1}{6}~\lambda\kappa^{4}_{5}~,
\end{equation}
\begin{equation}
\Lambda_{4}=\frac{1}{2}~\kappa^{2}_{5}\left(\Lambda_{5}+\frac{1}{6}~\kappa^{2}_{5}\lambda^{2}\right)
\end{equation}
and
\begin{equation}
\Pi_{\mu\nu}=-\frac{1}{4}~\tau_{\mu\alpha}\tau^{\alpha}_{\nu}+\frac{1}{12}~\tau\tau_{\mu\nu}+\frac{1}{8}~
q_{\mu\nu}\tau_{\alpha\beta}\tau^{\alpha\beta}-\frac{1}{24}~q_{\mu\nu}\tau^{2}
\end{equation}
and $E_{\mu\nu}$ is the electric part of the 5D Weyl tensor. Here
$\kappa_{5},~\Lambda_{5},~\tau_{\mu\nu}$ and $\Lambda_{4}$ are
respectively the 5D gravitational coupling constant, 5D
cosmological constant, the brane tension (vacuum energy), brane
energy-momentum tensor and effective 4D cosmological constant. The
explicit form of the above modified Einstein equations in flat
universe are

\begin{equation}
3H^{2}=\Lambda_{4}+\kappa^{2}_{4}\rho+\frac{\kappa^{2}_{4}}{2\lambda}~\rho^{2}+\frac{6}{\lambda
\kappa^{2}_{4}}\cal{U}
\end{equation}
and
\begin{equation}
2\dot{H}+3H^{2}=\Lambda_{4}-\kappa^{2}_{4}p-\frac{\kappa^{2}_{4}}{2\lambda}~\rho
p-\frac{\kappa^{2}_{4}}{2\lambda}~\rho^{2}-\frac{2}{\lambda
\kappa^{2}_{4}}\cal{U}
\end{equation}
The dark radiation $\cal{U}$ obeys

\begin{equation}
\dot{\cal U}+4H{\cal U}=0
\end{equation}
where $\rho=\rho_{nvmcg}+\rho_{m}$ and $p=p_{nvmcg}+p_{m}$ are the
total energy density and pressure respectively.

\noindent

As in the present problem the interaction between DE and
pressureless DM has been taken into account for interacting DE and
DM the energy balance equation will be
\begin{equation}
\dot{\rho}_{nvmcg}+3H\left(1+\omega_{nvmcg}\right)\rho_{nvmcg}=-Q,~~~for
~NVMCG~ and ~
\end{equation}
\begin{equation}
\dot{\rho}_m+3H\rho_m=Q, ~for~ the~ DM~ interacting ~with~ NVMCG.
\end{equation}
where $Q=3bH\rho$ is the interaction term, $b$ is the coupling
parameter (or transfer strength) and $\rho=\rho_{nvmcg}+\rho_m$ is
the total cosmic energy density which satisfies the energy
conservation equation $\dot{\rho}+3H\left(\rho+p\right)=0$
\cite{Guo1, del Campo1}.

\noindent

Since we lack information about the fact, how does DE and DM
interact so we are not able to estimate the interaction term from
the first principles. However, the negativity of $Q$ immediately
implies the possibility of having negative DE in the early
universe which is overruled by to the necessity of the second law
of thermodynamics to be held \cite{Alcaniz1}. Hence $Q$ must be
positive and small. From the observational data of 182 Gold type
Ia supernova samples, CMB data from the three year WMAP survey and
the baryonic acoustic oscillations from the Sloan Digital Sky
Survey, it is estimated that the coupling parameter between DM and
DE must be a small positive value (of the order of unity), which
satisfies the requirement for solving the cosmic coincidence
problem and the second law of thermodynamics \cite{Feng1}. Due to
the underlying interaction, the beginning of the accelerated
expansion is shifted to higher redshifts. The continuity equations
for dark energy and dark matter are given in equations (9) and
(10). Now we shall study the dynamical system assuming
$\Lambda_{4}={\cal U}=0$ (in absence of cosmological constant and
dark radiation).

\subsection{DYNAMICAL SYSTEM ANALYSIS}

\noindent

In this subsection we plan to analyze the dynamical system. For
that firstly we convert the physical parameters into some
dimensionless form, given by
\begin{equation}\label{7}
x=\ln a, ~~ u=\frac{\rho_{nvmcg}}{3H^2}, ~~
v=\frac{\rho_m}{3H^2},~~ y=\frac{a}{3H^{2}}
\end{equation}
where the present value of the scale factor $a_0=1$ is assumed.
Using eqns. (1), (6), (7), (9) and (10) into (\ref{7}) we get the
parameter gradients as

\begin{equation}\label{8}
\frac{du}{dx}=-3b\left(u+v\right)-3\left(u+u\omega_{nvmcg}^{(RSII)}\right)-6\frac{\dot{H}}{X}u
\end{equation}
,
\begin{equation}\label{9}
\frac{dv}{dx}=3b(u+v)-3v-6\frac{\dot{H}}{X}v
\end{equation}
and

\begin{equation}
\frac{dy}{dx}=y\left(1-6\frac{\dot{H}}{X}\right)
\end{equation}

where $\omega_{nvmcg}^{(RSII)}$ is the EoS parameter for NVMCG
determined as
\begin{equation}\label{10}
\omega_{nvmcg}^{(RSII)}=\frac{p_{nvmcg}}{\rho_{nvmcg}}=\frac{1}{y^{n}X^{n}}\left(A-\frac{By^{n-m}}{u^{\alpha+1}X^{m-n+\alpha+1}}\right),
\end{equation}

\begin{equation}
\dot{H}=\frac{\lambda}{u+v}\left(\kappa^{2}-\frac{1}{u+v}\right)\left[u\omega_{nvmcg}^{(RSII)}+\left\{\left(\frac{1}{\kappa^{2}\left(u+v\right)}-1\right)\left(u\left(\omega_{nvmcg}^{RSII}+2\right)+2v\right)\right\}\right]
\end{equation}
and
\begin{equation}
X=\frac{2\lambda}{u+v}\left[\frac{1}{\kappa^{2}\left(u+v\right)}-1\right]
\end{equation}

\subsubsection{\bf CRITICAL POINTS}

\noindent

The critical points of the above system are obtained by putting
$\frac{du}{dx}=\frac{dv}{dx}=\frac{dy}{dx}=0$. But due to the
complexity of these equations, it is not possible to find a
solution in terms of all the involved parameters. So we find a
solution for the above system, by putting the following numerical
values to some of the parameters appearing in the system. We take,
$$\alpha=0.5, ~~b=0.5,~~ n=1,~~ m=1$$ and obtain the following
critical point,
\begin{equation}
u_{c}=\frac{0.3125}{\kappa^{2}},~~~~~~
v_{c}=\frac{0.1875}{\kappa^{2}},~~~~~~
~~~y_{c}=\frac{1.02337\times10^{-20}}{\lambda^{5}}\left(\frac{8.19379\times10^{18}B\lambda^{\frac{5}{2}}}{\kappa^{2}}-\frac{1.14512\times10^{19}A\lambda^{4}}{\kappa^{2}}\right)
\end{equation}

\noindent

The critical point correspond to the era dominated by DM and NVMCG
type DE. For the critical point $(u_{c},v_{c})$, the equation of
state parameter given by equation (15) of the interacting DE takes
the form

\begin{equation}
\omega_{nvmcg}^{(RSII)}=\frac{p_{nvmcg}}{\rho_{nvmcg}}=\frac{1}{y_{c}^{n}X^{n}}\left(A-\frac{By_{c}^{n-m}}{u_{c}^{\alpha+1}X^{m-n+\alpha+1}}\right),
\end{equation}
where
\begin{equation}
X=\frac{2\lambda}{u_{c}+v_{c}}\left[\frac{1}{\kappa^{2}\left(u_{c}+v_{c}\right)}-1\right]
\end{equation}

\vspace{2mm}

\begin{figure}
\includegraphics[height=3in]{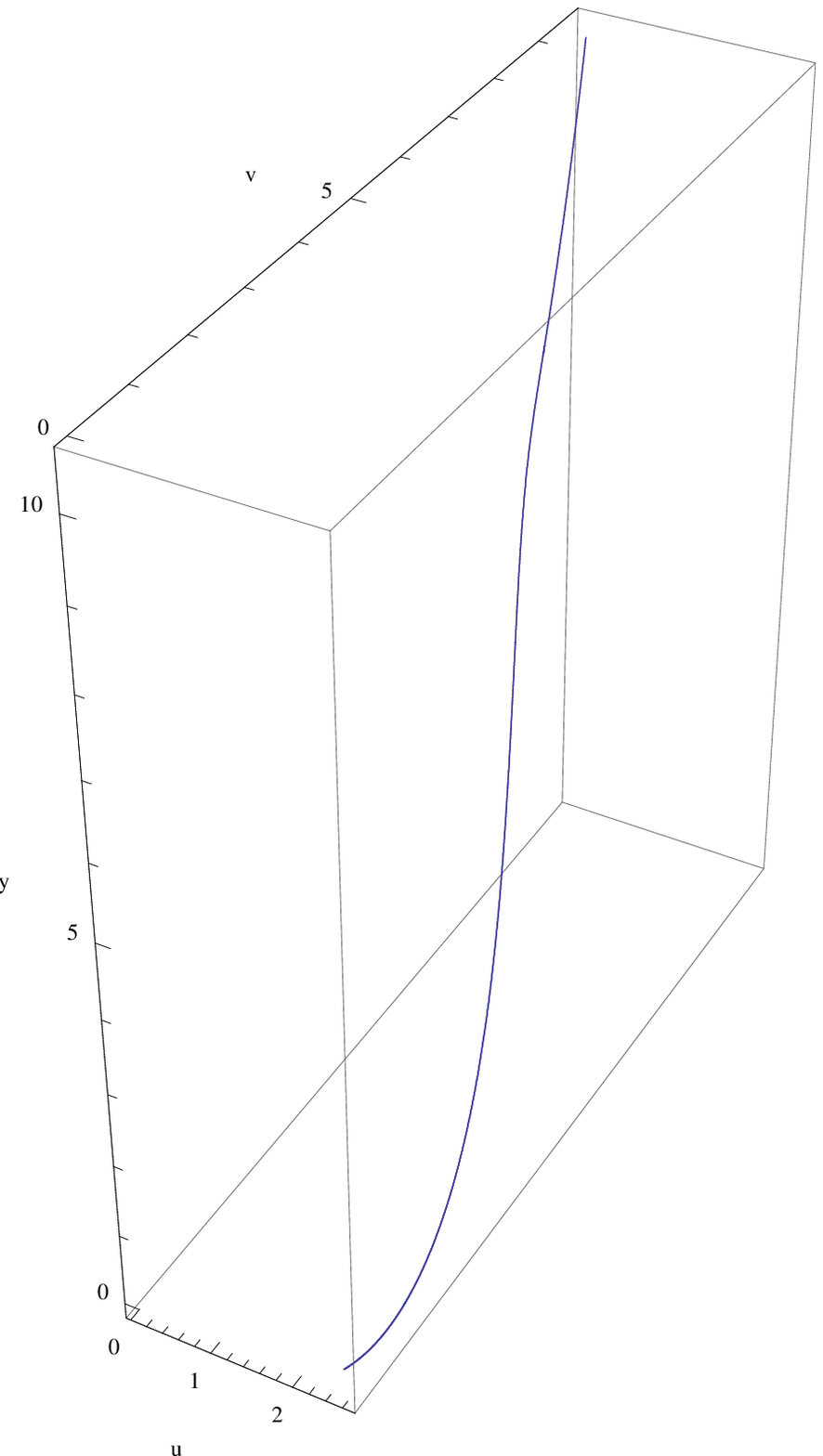}~~~~\\

~~~~~~~~~~~~~~~~~~~~~~~~~~~~~~~~~~~~~Fig.1~~~~~~~~~~~~~~~~~~~~~~~~~~~~~~~~~\\
\vspace{1cm}
\end{figure}

Fig 1 : The dimensionless density parameters $u$, $v$ and $y$ are
plotted against each other in a 3D-scenario. Other parameters are
fixed at $\alpha=0.5, b=0.5, A_{0}=1/3, B_{0}=3, n=2, m=1,
\lambda=1.5,$ and $\kappa=0.2$.\\

\begin{figure}
\includegraphics[scale=0.28]{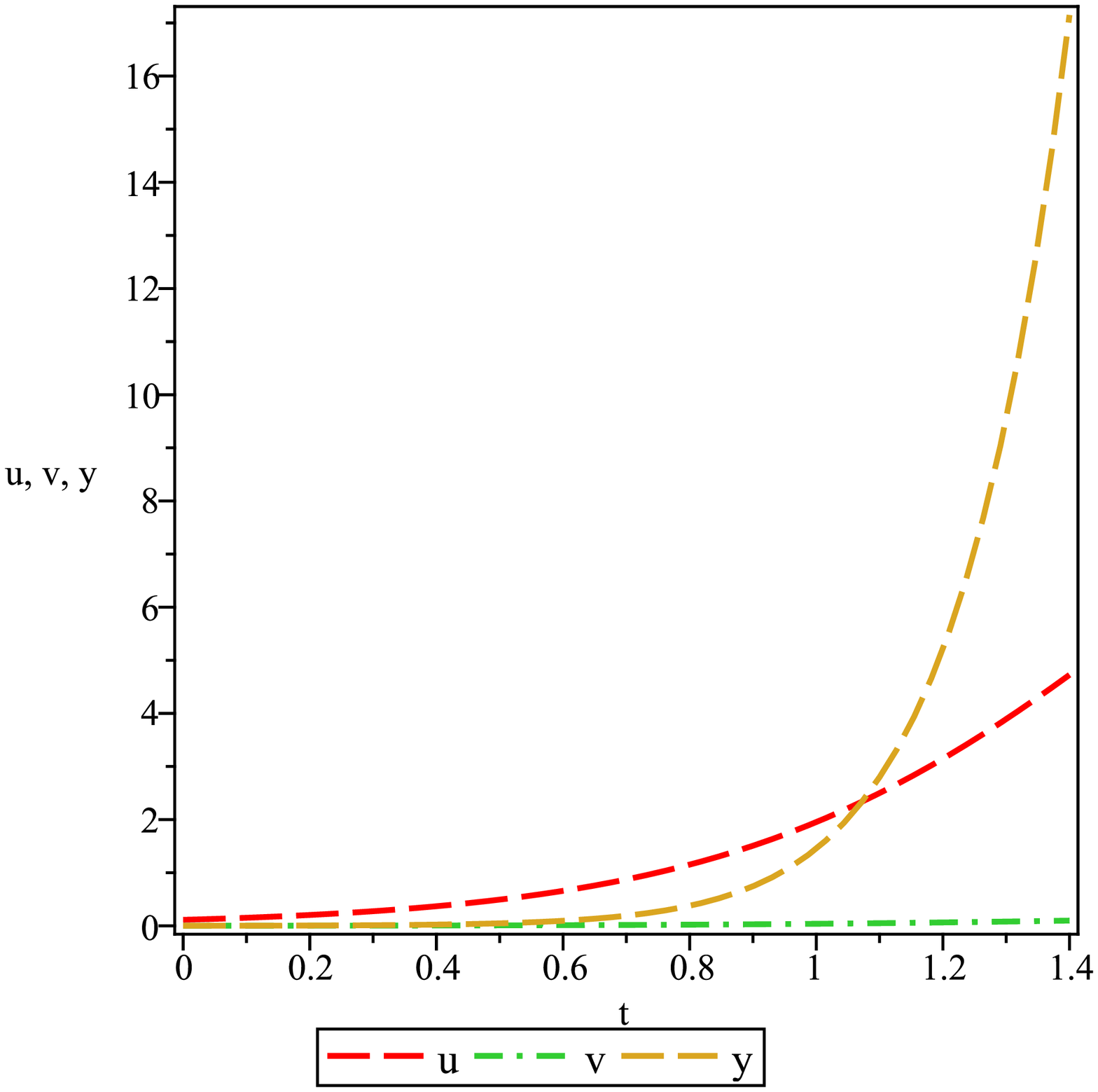}~~~~~~~\includegraphics[scale=0.28]{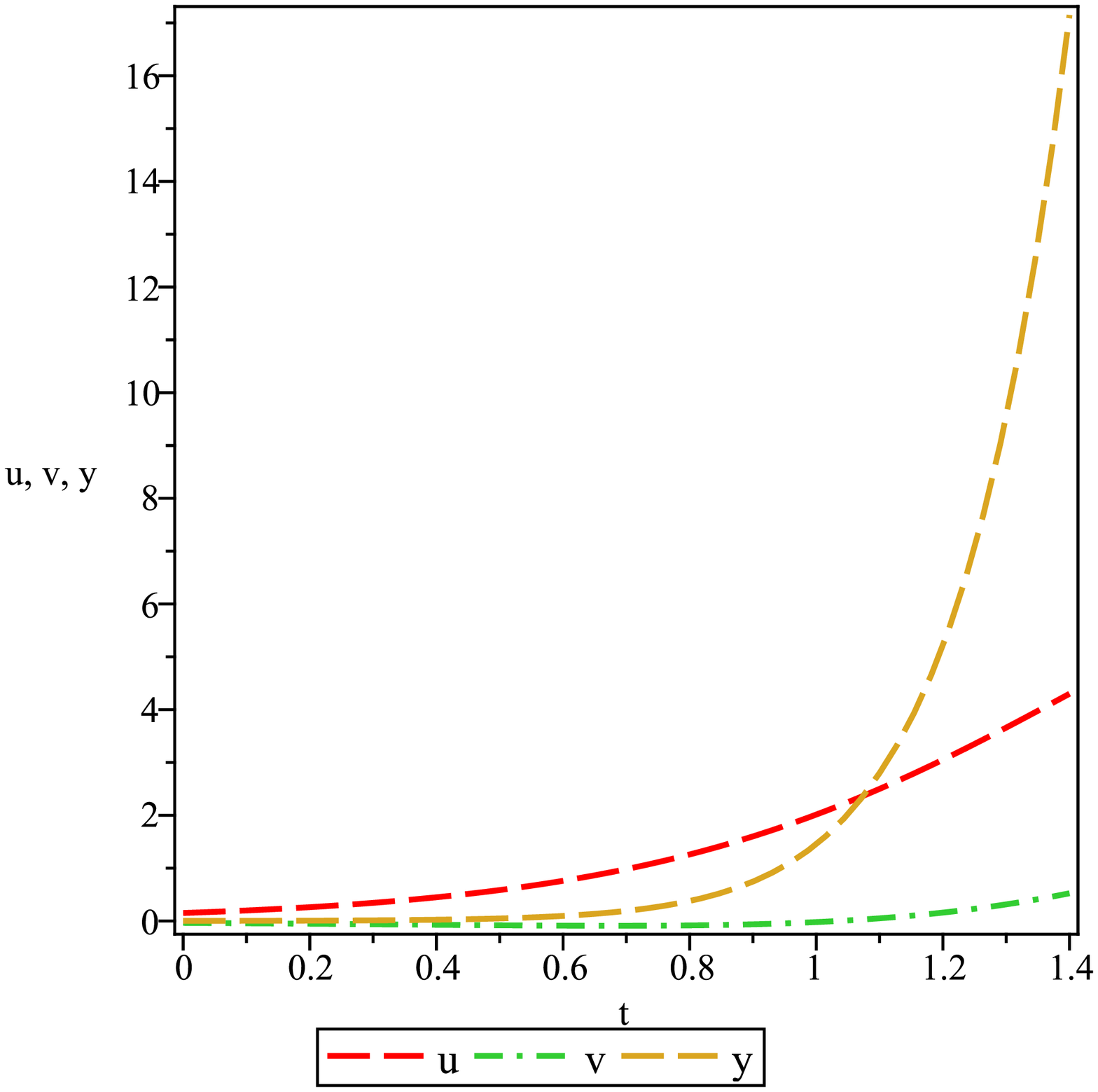}~~~~~\includegraphics[scale=0.28]{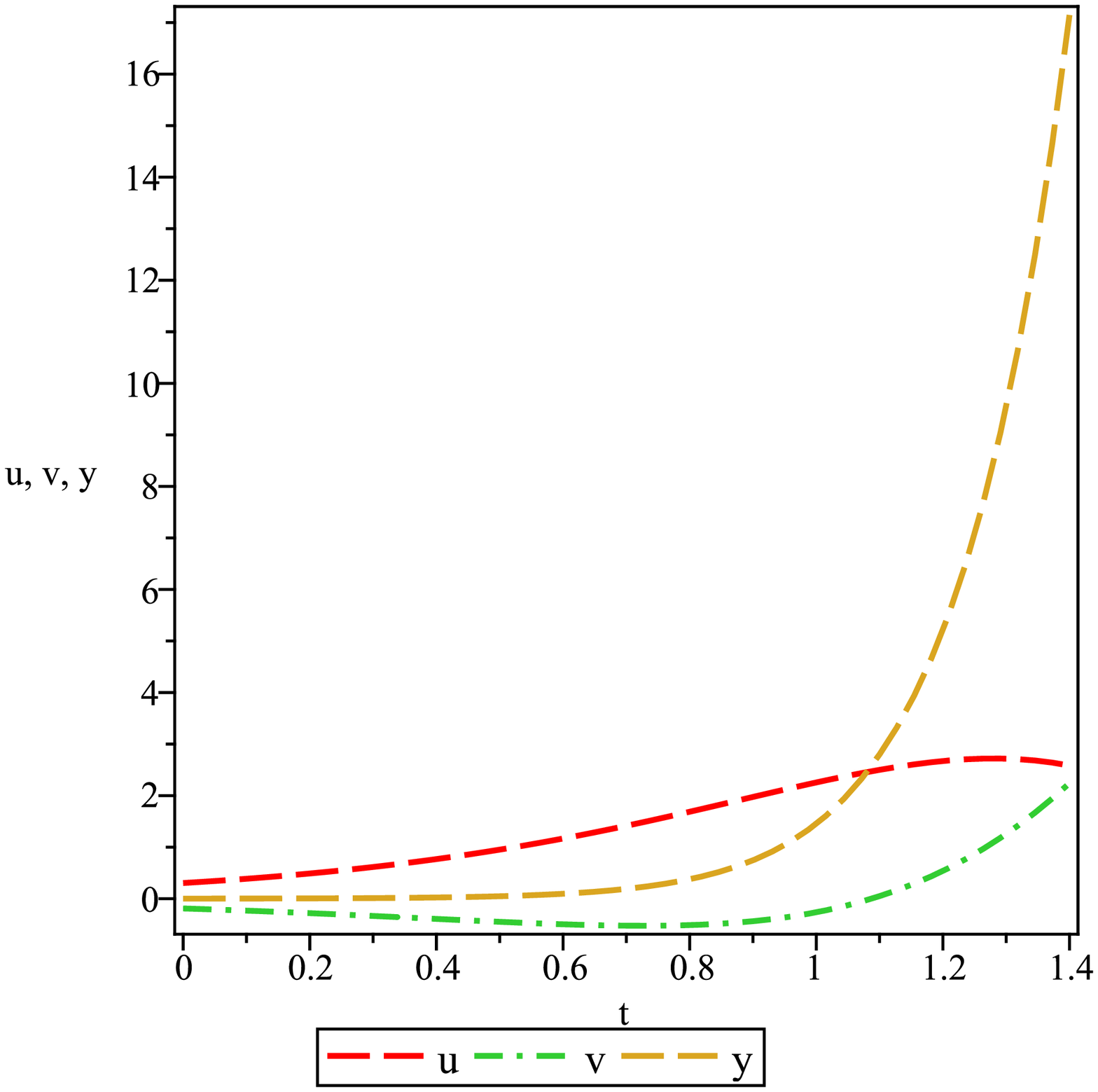}~

~~~~~~~~~~~~~~~~~~~~Fig.2~~~~~~~~~~~~~~~~~~~~~~~~~~~~~~~~~~~~~~~~~~~~~~~~~~~~~~~~~~~~~~Fig.3~~~~~~~~~~~~~~~~~~~~~~~~~~~~~~~~~~~~~~~~~~~~~~~~~~~~~~~~~~~~~~~~~~~Fig.4~~~~~~~~~~~~~~~~~~~~~~~~~~~~~~~~~\\
\vspace{1cm}
\end{figure}

Fig 2 : The dimensionless density parameters are plotted against
e-folding time. The initial conditions are $v(1.1)=0.05,
u(1.1)=2.5$ and $y(1.1)=2.8$. Other parameters are fixed at
$\alpha=0.5, b=0.001, A_{0}=1/3, B_{0}=3, n=5, m=1, \lambda=1.5,$
and $\kappa=0.2$.\\
Fig 3 : The dimensionless density parameters are plotted against
e-folding time. The initial conditions are $v(1.1)=0.05,
u(1.1)=2.5$ and $y(1.1)=2.8$. Other parameters are fixed at
$\alpha=0.5, b=0.1, A_{0}=1/3, B_{0}=3, n=5, m=1, \lambda=1.5,$
and $\kappa=0.2$.\\
\\Fig 4 :  The dimensionless density parameters are plotted against
e-folding time. The initial conditions are $v(1.1)=0.05,
u(1.1)=2.5$ and $y(1.1)=2.8$. Other parameters are fixed at
$\alpha=0.5, b=0.5, A_{0}=1/3, B_{0}=3, n=5, m=1, \lambda=1.5,$
and $\kappa=0.2$.\\\\\\\\

\begin{figure}
\includegraphics[scale=0.28]{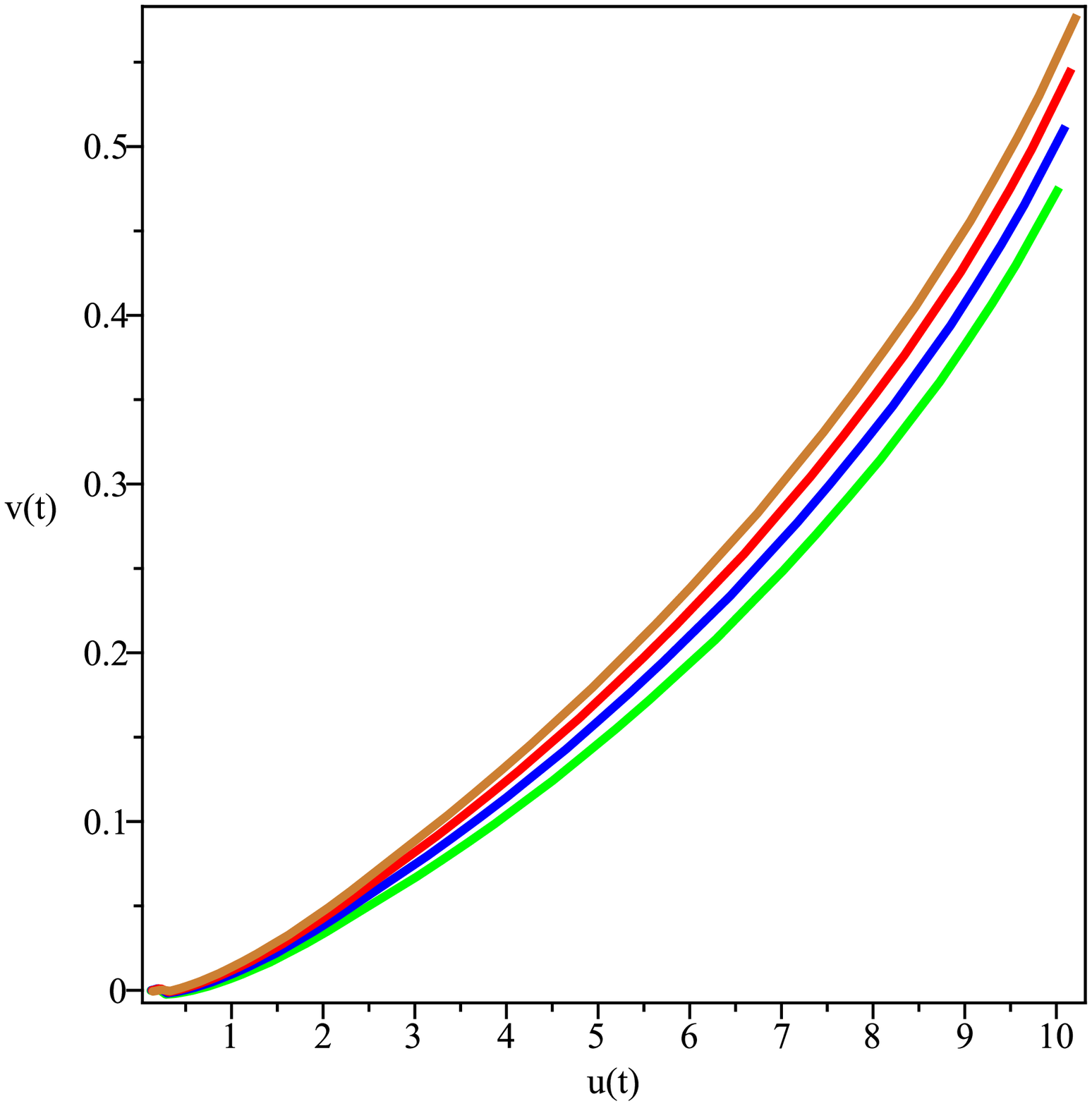}~~~~~~~~\includegraphics[scale=0.28]{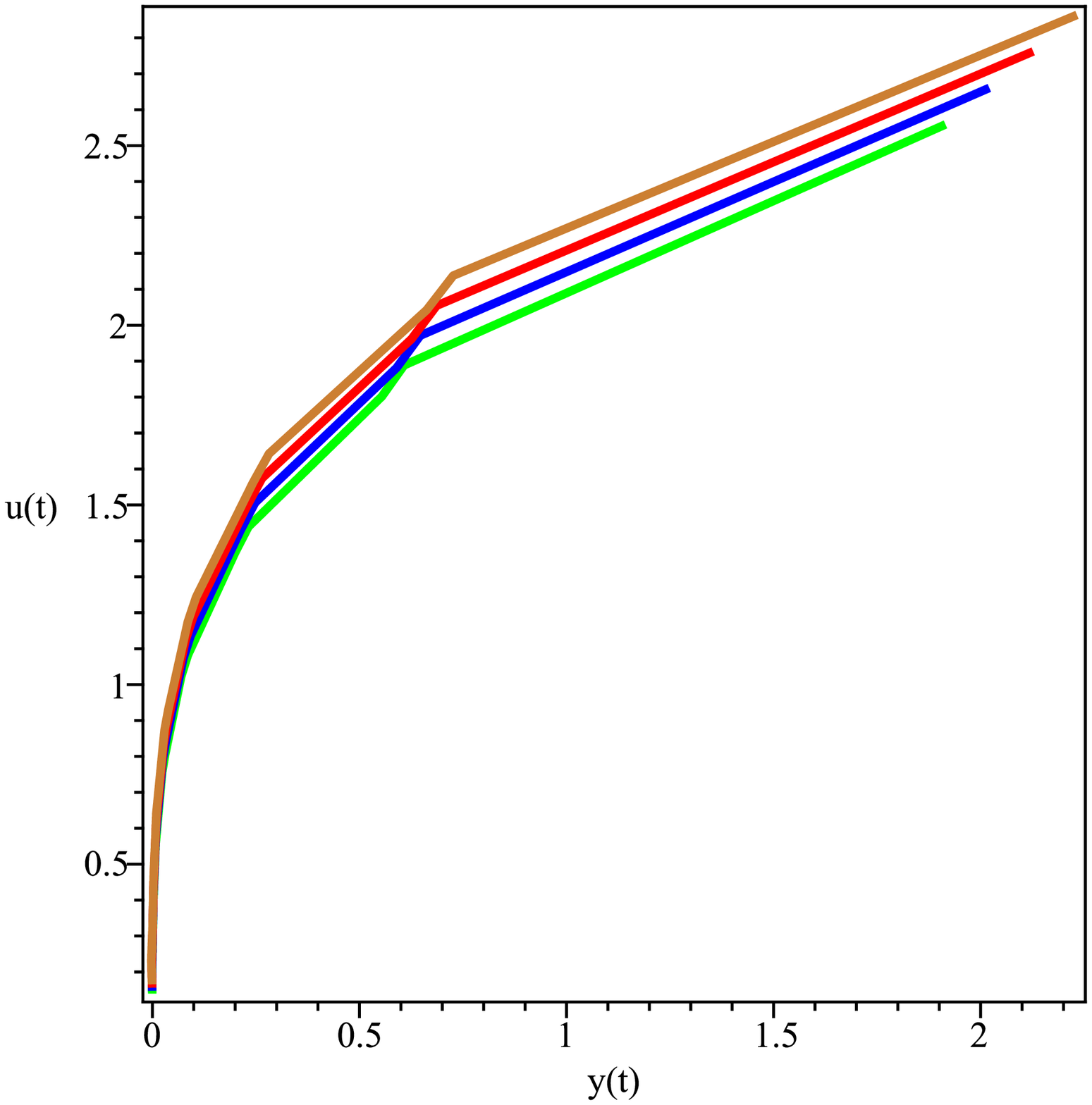}~~~~~~~~\includegraphics[scale=0.28]{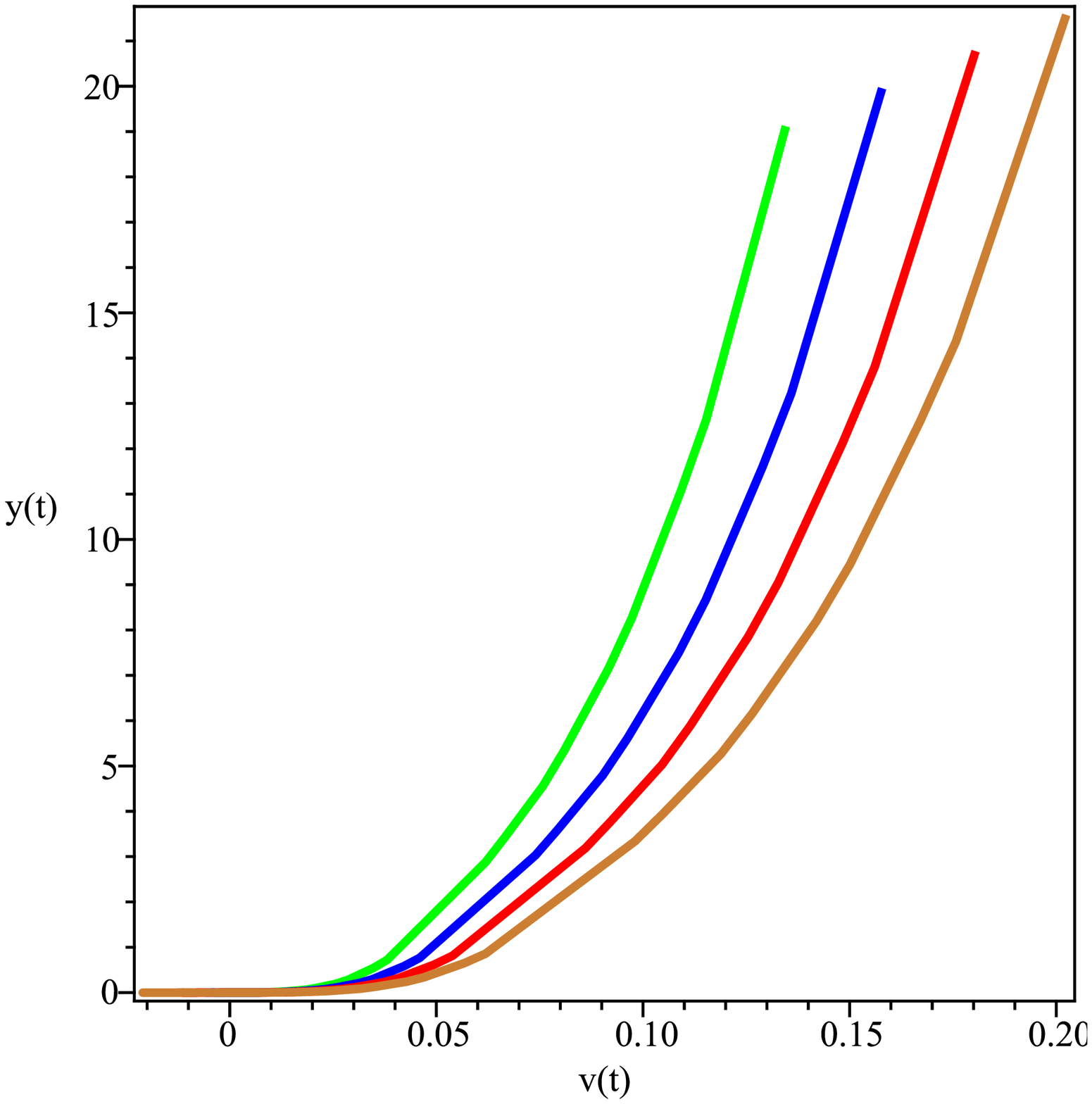}~

~~~~~~~~~~~~~~Fig.5~~~~~~~~~~~~~~~~~~~~~~~~~~~~~~~~~~~~~~~~~~~~~~Fig. 6~~~~~~~~~~~~~~~~~~~~~~~~~~~~~~~~~~~~~~~~Fig. 7~~~~~~~~~~~~~~~~~~~~~~~~~~~\\
\vspace{1cm}
\end{figure}

Fig 5 :  The phase diagram of the parameters $u(t)$ and $v(t)$
depicting an attractor solution. The initial conditions chosen are
$v(1)=0.05, u(0)=2.5, y(1)=1.8$ (green); $v(1)=0.06, u(1)=2.6,
y(1)=1.9$ (blue); $v(1)=0.07, u(1)=2.7, y(1)=2$ (red); $v(1)=0.08,
u(1)=2.8, y(1)=2.1$ (gold). Other
parameters are fixed at $\alpha=0.5, b=0.01, A=1/3, B=3, n=5, m=1, \lambda=1.5$ and $\kappa=0.2$.\\
\\Fig 6 :  The dimensionless density parameters $u(t)$ and $y(t)$ are plotted against
e-folding time. The initial conditions are $v(1.1)=0.05,
u(1.1)=2.5$ and $y(1.1)=2.8$. Other parameters are fixed at
$\alpha=0.5, b=0.01, A_{0}=1/3, B_{0}=3, n=5, m=1, \lambda=1.5,$
and $\kappa=0.2$.\\
Fig 7 :  The dimensionless density parameters $v(t)$ and $y(t)$
are plotted against e-folding time. The initial conditions are
$v(1.1)=0.05, u(1.1)=2.5$ and $y(1.1)=2.8$. Other parameters are
fixed at $\alpha=0.5, b=0.001, A_{0}=1/3, B_{0}=3, n=5, m=1,
\lambda=1.5,$
and $\kappa=0.2$.\\

\begin{figure}
\includegraphics[height=3in]{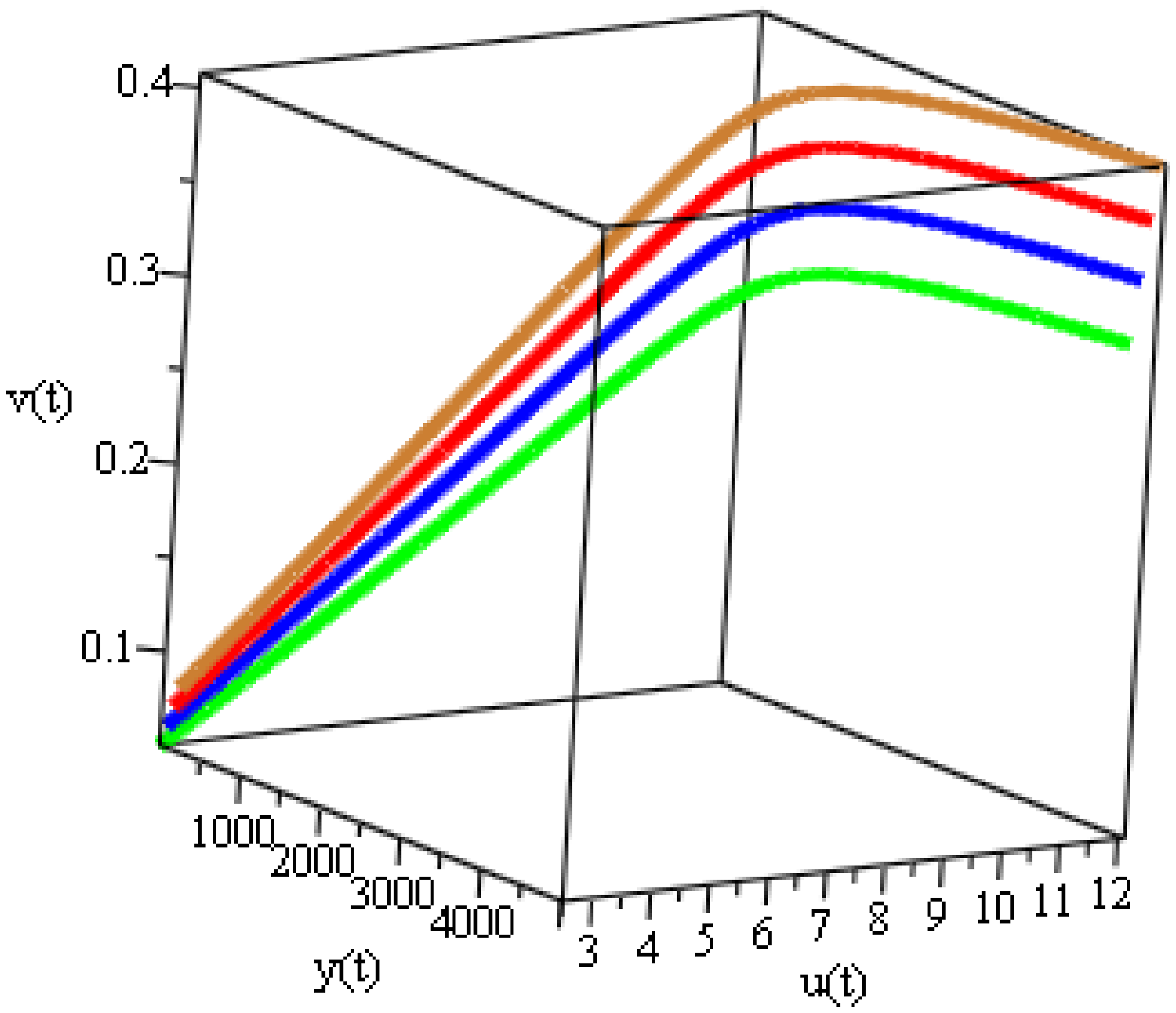}~~~~\\

~~~Fig.8~~~~~~~~~~~~~~~~~~~~~~~~~~~~~~~~~\\
\vspace{1cm}
\end{figure}

Fig 8 :  A 3D-phase portrait of the dimensionless density
parameters $u(t)$, $v(t)$ and $y(t)$ is plotted against each
other. The initial conditions are $v(1.1)=0.05, u(1.1)=2.5$ and
$y(1.1)=2.8$. Other parameters are fixed at $\alpha=0.5, b=0.001,
A_{0}=1/3, B_{0}=3, n=5, m=1, \lambda=1.5,$
and $\kappa=0.2$.\\

\begin{figure}
\includegraphics[height=3in]{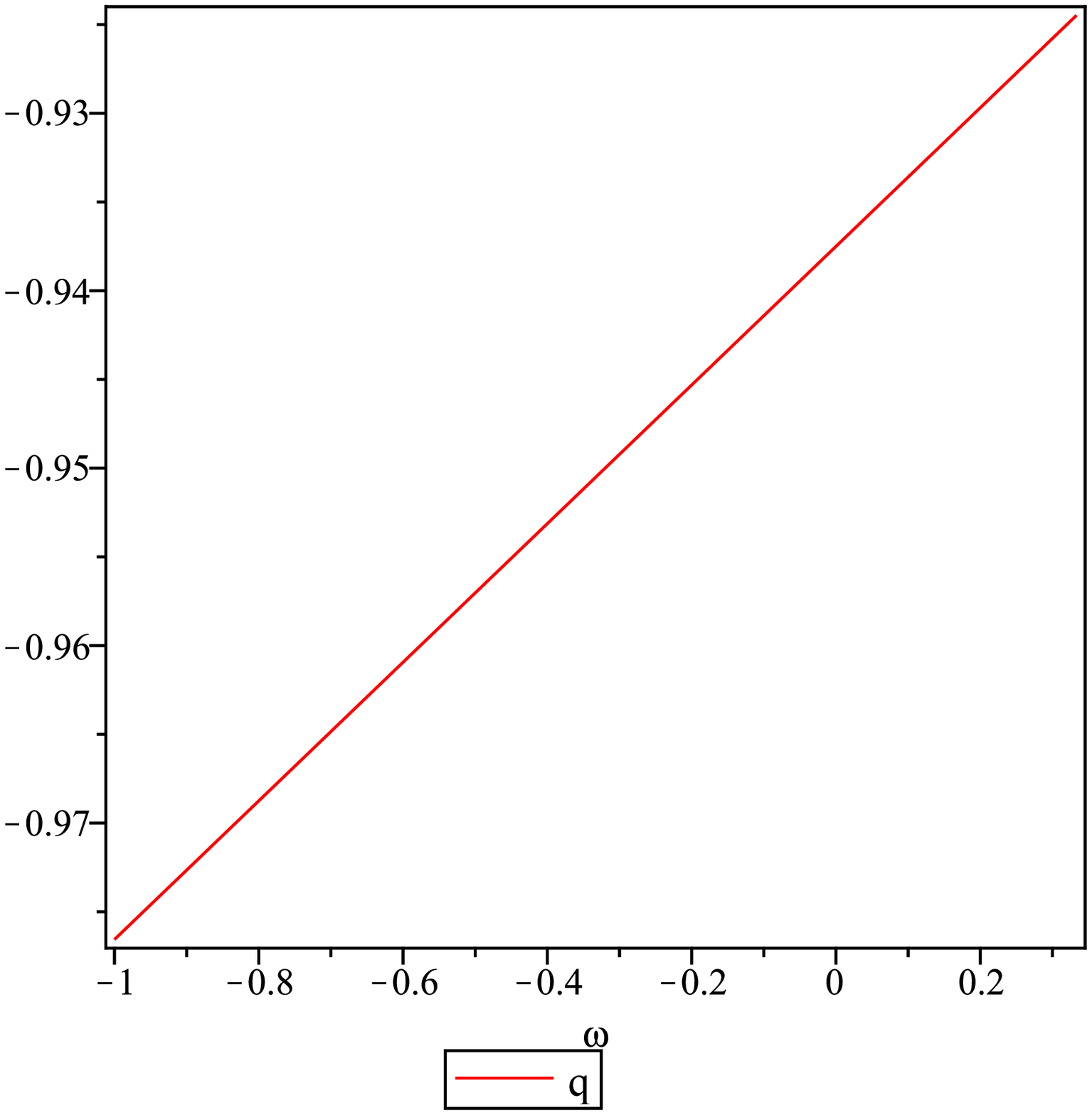}~~~~\\

~~Fig.9~~~~~~~~~~~~~~~~~~~~~~~~~~~~~~~~~\\
\vspace{1cm}
\end{figure}

Fig 9 :   The deceleration parameter is plotted against the EoS
parameter. Other parameters are fixed at $\alpha=0.5, b=0.001,
A_{0}=1/3, B_{0}=3, n=5, m=1, \lambda=1.5,$
and $\kappa=0.2$.\\

\begin{figure}
\includegraphics[height=3in]{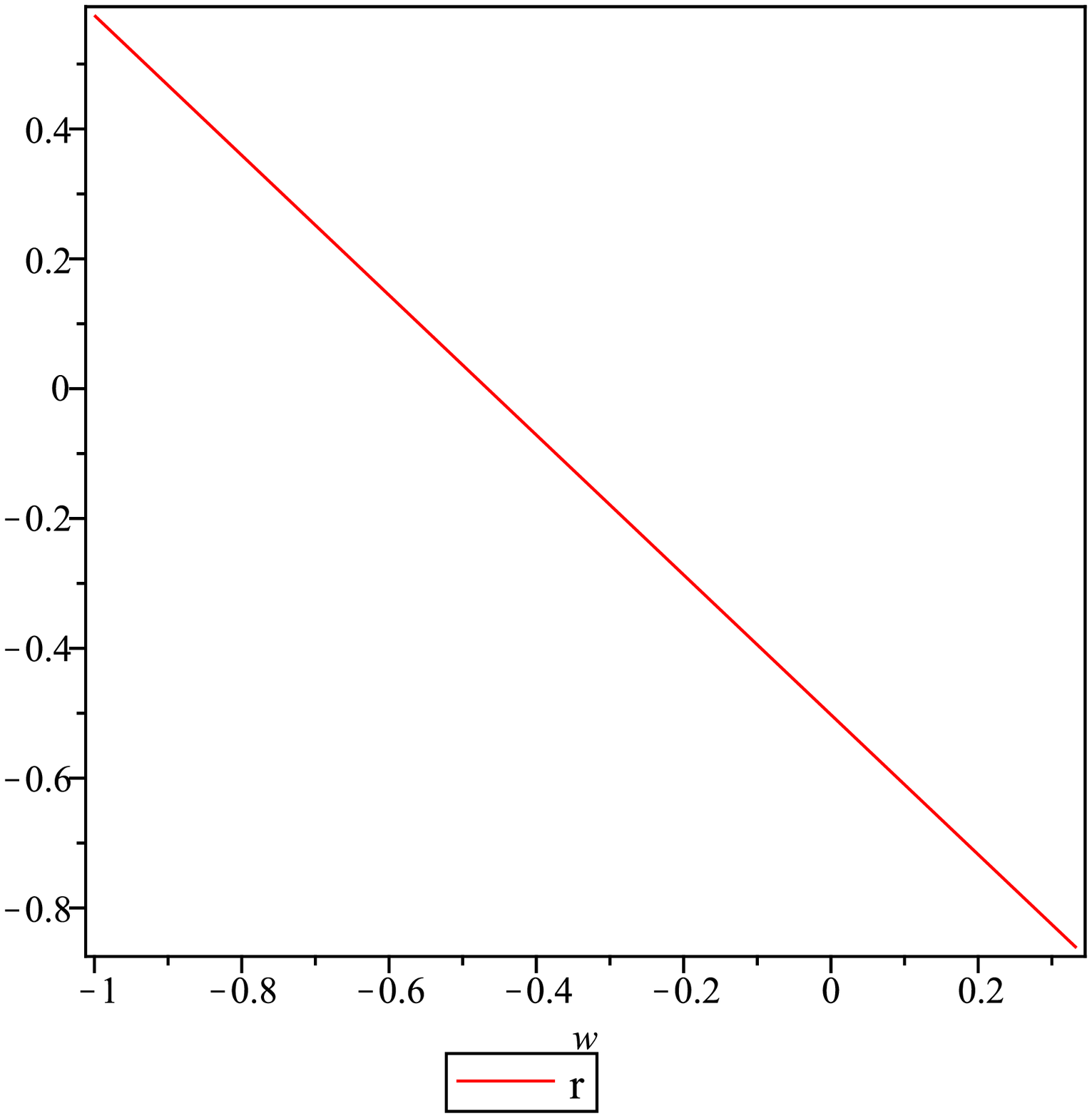}~~~~\\

~~Fig.10~~~~~~~~~~~~~~~~~~~~~~~~~~~~~~~~~\\
\vspace{1cm}
\end{figure}
Fig 10 :  The statefinder parameter $r$ is plotted against the EoS
parameter. Other parameters are fixed at $\alpha=0.5, b=0.001,
A_{0}=1/3, B_{0}=3, n=5, m=1, \lambda=1.5,$ and $\kappa=0.2$.
\\
\begin{figure}
\includegraphics[height=3in]{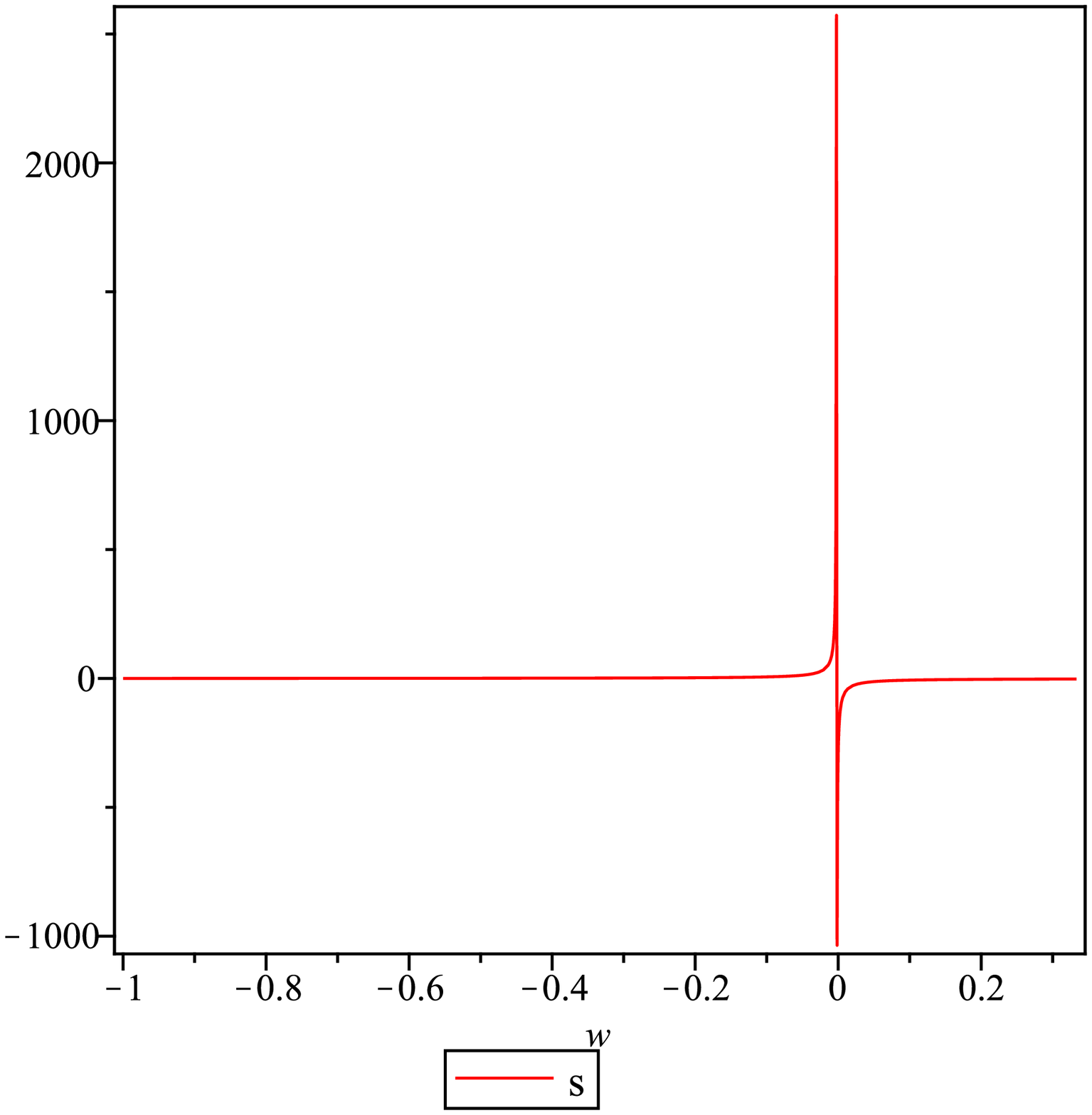}~~~~\\

~~Fig.11~~~~~~~~~~~~~~~~~~~~~~~~~~~~~~~~~\\
\vspace{1cm}
\end{figure}
Fig 11 : The statefinder parameter $s$ is plotted against the EoS
parameter. Other parameters are fixed at $\alpha=0.5, b=0.001,
A_{0}=1/3, B_{0}=3, n=5, m=1, \lambda=1.5,$ and $\kappa=0.2$.
\\
\begin{figure}
\includegraphics[height=3in]{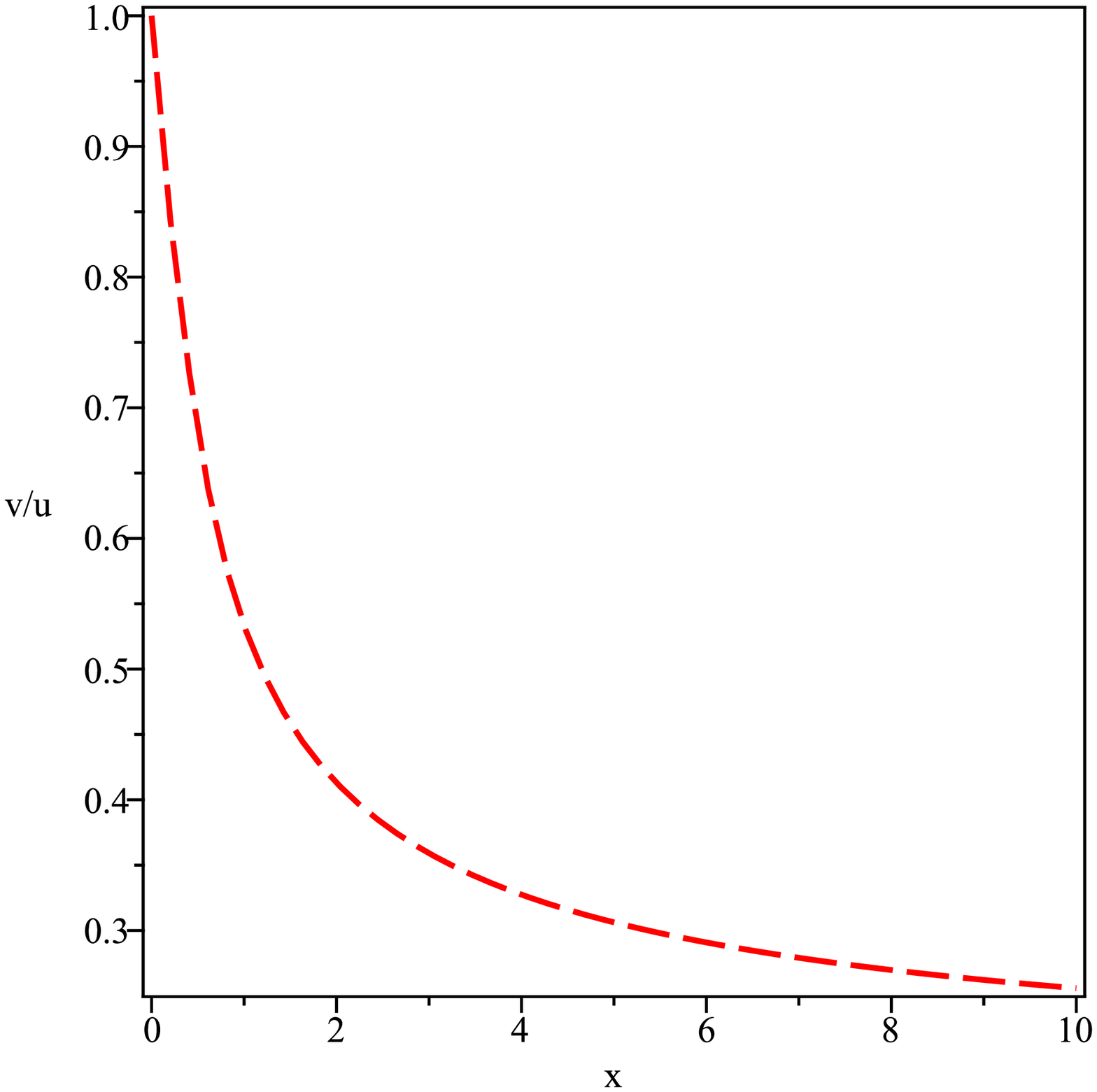}~~~~\\

~~Fig.12~~~~~~~~~~~~~~~~~~~~~~~~~~~~~~~~~\\
\vspace{1cm}
\end{figure}
Fig 12 : The ratio of density parameters is shown against
e-folding time. The initial conditions chosen are v(1)=0.05,
u(1)=2.5, y(1)=1.8. Other parameters are fixed at $\alpha=0.5,
b=0.001, A_{0}=1/3, B_{0}=3, n=5, m=1, \lambda=1.5,$ and $\kappa=0.2$.\\

\subsubsection{\bf STABILITY AROUND CRITICAL POINT}

Now we check the stability of the dynamical system  (eqs. (12) and
(13) and (14)) about the critical point. In order to do this, we
linearize the governing equations about the critical point i.e.,
\begin{equation}
u=u_c+\delta u ,~~~~   v=v_c+\delta v,~~~ ~~~ y=y_{c}+\delta y
\end{equation}

Now if we assume $f=\frac{du}{dx}$ , $g=\frac{dv}{dx}$ and
$h=\frac{dy}{dx}$ then we may obtain
\begin{equation}
\delta\left(\frac{du}{dx}\right)=\left[\partial_{u}
f\right]_{c}\delta u+\left[\partial_{v} f\right]_{c}\delta
v+\left[\partial_{y} f\right]_{c}\delta y
\end{equation}

\begin{equation}
\delta\left(\frac{dv}{dx}\right)=\left[\partial_{u}
g\right]_{c}\delta u+\left[\partial_{v} g\right]_{c}\delta
v+\left[\partial_{y} g\right]_{c}\delta y
\end{equation}
and

\begin{equation}
\delta\left(\frac{dy}{dx}\right)=\left[\partial_{u}
h\right]_{c}\delta u+\left[\partial_{v} h\right]_{c}\delta
v+\left[\partial_{y} h\right]_{c}\delta y
\end{equation}

\vspace{2mm}

where

\vspace{2mm}

\noindent
\begin{eqnarray*}
\partial_{u}{f}=-3\frac{2^{-1-m-n-\alpha}u^{-1-\alpha}y^{-m-n}}{\lambda(u+v)^{2}(-1+(u+v)\kappa^{2})^{2}}
\left(\frac{\lambda(-u-v+\frac{1}{\kappa^{2}})}{(u+v)^{2}}\right)^{-m-n-\alpha}
\end{eqnarray*}
\begin{eqnarray*}
\times\left[-2^{n}Bv(u+v)^{2}y^{n}\kappa^{2}(u+2mu+u\alpha-v
\alpha-(u+v)(mu-v\alpha)\kappa^{2})\left(\frac{\lambda(-u-v+
\frac{1}{\kappa^{2}})}{(u+v)^{2}}\right)^{n}+2^{1+m+\alpha}u^{\alpha}y^{m}\right.
\end{eqnarray*}
\begin{eqnarray*}
\left.\times(-1+(u+v)\kappa^{2})\lambda
\left(\frac{\lambda(-u-v+\frac{1}{\kappa^{2}})}{(u+v)^{2}}\right)
^{m+\alpha}\left(Av(-2nu-v+(u+v)(nu+v)\kappa^{2})+2^{n}(u+v)^{2}y^{n}
\right.\right.
\end{eqnarray*}
\begin{equation}
\left.\left.\times(-1+(u+v)\kappa^{2})(-1+b+2(2u+v)\kappa^{2})\left(\frac{
\lambda(-u-v+\frac{1}{\kappa^{2}})}{(u+v)^{2}}\right)^{-m-n-\alpha}\right)\right]
\end{equation}

\vspace{2mm}

\noindent

\begin{eqnarray*}
\partial_{v}{f}=-3\frac{2^{-1-m-n-\alpha}u^{-\alpha}y^{-m-n}}{\lambda(u+v)^{2}(-1+(u+v)\kappa^{2})^{2}}
\left(\frac{\lambda(-u-v+\frac{1}{\kappa^{2}})}{(u+v)^{2}}\right)^{-m-n-\alpha}
\end{eqnarray*}
\begin{eqnarray*}
\times\left[2^{n}B(u+v)^{2}y^{n}\kappa^{2}(-u-2v(1+m+\alpha)+(u+v)(u+v(1+m+\alpha))\kappa^{2})\left(\frac{\lambda(-u-v+
\frac{1}{\kappa^{2}})}{(u+v)^{2}}\right)^{n}+2^{1+m+\alpha}u^{\alpha}\right.
\end{eqnarray*}
\begin{eqnarray*}
\left.\times y^{m}(-1+(u+v)\kappa^{2})\lambda
\left(\frac{\lambda(-u-v+\frac{1}{\kappa^{2}})}{(u+v)^{2}}\right)
^{m+\alpha}\left(Av(-u-2nv+(u+v)(u+nv)\kappa^{2})+2^{n}(u+v)^{2}y^{n}
\right.\right.
\end{eqnarray*}
\begin{equation}
\left.\left.\times(b+2u\kappa^{2})(-1+(u+v)\kappa^{2})\left(\frac{
\lambda(-u-v+\frac{1}{\kappa^{2}})}{(u+v)^{2}}\right)^{n}\right)\right]
\end{equation}

\vspace{2mm}

\noindent

\begin{equation}
\partial_{y}{f}=\frac{2^{-n}3Anuvy^{-1-n}
\left(\frac{\lambda(-u-v+\frac{1}{\kappa^{2}})}{(u+v)^{2}}\right)^{-n}}{u+v}
+\frac{2^{-1-m-\alpha}3Bmu^{-\alpha}v(u+v)y^{-1-m}\kappa^{2}
\left(\frac{\lambda(-u-v+\frac{1}{\kappa^{2}})}{(u+v)^{2}}\right)^{-m-\alpha}}
{(-1+(u+v)\kappa^{2})\lambda}
\end{equation}

\vspace{2mm}

\noindent

\begin{eqnarray*}
\partial_{u}{g}=3\frac{2^{-1-m-n-\alpha}u^{-1-\alpha}y^{-m-n}}{\lambda(u+v)^{2}(-1+(u+v)\kappa^{2})^{2}}
\left(\frac{\lambda(-u-v+\frac{1}{\kappa^{2}})}{(u+v)^{2}}\right)^{-m-n-\alpha}
\end{eqnarray*}
\begin{eqnarray*}
\times\left[-2^{n}Bv(u+v)^{2}y^{n}\kappa^{2}(u+2mu+u\alpha-v
\alpha-(u+v)(mu-v\alpha)\kappa^{2})\left(\frac{\lambda(-u-v+
\frac{1}{\kappa^{2}})}{(u+v)^{2}}\right)^{n}+2^{1+m+\alpha}u^{1+\alpha}y^{m}\right.
\end{eqnarray*}
\begin{eqnarray*}
\left.\times(-1+(u+v)\kappa^{2})\lambda
\left(\frac{\lambda(-u-v+\frac{1}{\kappa^{2}})}{(u+v)^{2}}\right)
^{m+\alpha}\left(Av(-2nu-v+(u+v)(nu+v)\kappa^{2})+2^{n}(u+v)^{2}y^{n}
\right.\right.
\end{eqnarray*}
\begin{equation}
\left.\left.\times(b-2v\kappa^{2})(-1+(u+v)\kappa^{2})\left(\frac{
\lambda(-u-v+\frac{1}{\kappa^{2}})}{(u+v)^{2}}\right)^{n}\right)\right]
\end{equation}

\vspace{2mm}

\noindent

\begin{eqnarray*}
\partial_{v}{g}=3\frac{2^{-1-m-n-\alpha}u^{-\alpha}y^{-m-n}}{\lambda(u+v)^{2}(-1+(u+v)\kappa^{2})^{2}}
\left(\frac{\lambda(-u-v+\frac{1}{\kappa^{2}})}{(u+v)^{2}}\right)^{-m-n-\alpha}
\end{eqnarray*}
\begin{eqnarray*}
\times\left[2^{n}B(u+v)^{2}y^{n}\kappa^{2}(-u-2v(1+m+\alpha)+(u+v)(u+v(1+m+\alpha))\kappa^{2})\left(\frac{\lambda(-u-v+
\frac{1}{\kappa^{2}})}{(u+v)^{2}}\right)^{n}-2^{1+m+\alpha}u^{\alpha}\right.
\end{eqnarray*}
\begin{eqnarray*}
\left.\times y^{m}(-1+(u+v)\kappa^{2})\lambda
\left(\frac{\lambda(-u-v+\frac{1}{\kappa^{2}})}{(u+v)^{2}}\right)
^{m+\alpha}\left(Au(u+2nv-(u+v)(u+nv)\kappa^{2})+2^{n}(u+v)^{2}y^{n}
\right.\right.
\end{eqnarray*}
\begin{equation}
\left.\left.\times(-1-b+2(u+2v)\kappa^{2})(-1+(u+v)\kappa^{2})\left(\frac{
\lambda(-u-v+\frac{1}{\kappa^{2}})}{(u+v)^{2}}\right)^{n}\right)\right]
\end{equation}

\vspace{2mm}

\noindent

\begin{equation}
\partial_{y}{g}=\frac{2^{-n}3Anuvy^{-1-n}
\left(\frac{\lambda(-u-v+\frac{1}{\kappa^{2}})}{(u+v)^{2}}\right)^{-n}}{u+v}
-\frac{2^{-1-m-\alpha}Bmu^{-\alpha}v(u+v)y^{-1-m}\kappa^{2}
\left(\frac{\lambda(-u-v+\frac{1}{\kappa^{2}})}{(u+v)^{2}}\right)^{-m-\alpha}}
{(-1+(u+v)\kappa^{2})\lambda}
\end{equation}

\vspace{2mm}

\noindent

\begin{eqnarray*}
\partial_{u}{h}=-3\frac{2^{-1-m-n-\alpha}u^{-1-\alpha}y^{-m-n}}{\lambda(u+v)^{2}(-1+(u+v)\kappa^{2})^{2}}
\left(\frac{\lambda(-u-v+\frac{1}{\kappa^{2}})}{(u+v)^{2}}\right)^{-m-n-\alpha}
\end{eqnarray*}
\begin{eqnarray*}
\times\left[-2^{n}B(u+v)^{2}y^{n}\kappa^{2}(-u-2mu-u\alpha+v
\alpha+(u+v)(mu-v\alpha)\kappa^{2})\left(\frac{\lambda(-u-v+
\frac{1}{\kappa^{2}})}{(u+v)^{2}}\right)^{n}+2^{1+m+\alpha}u^{1+\alpha}y^{m}\right.
\end{eqnarray*}
\begin{eqnarray*}
\left.\times(-1+(u+v)\kappa^{2})\lambda
\left(\frac{\lambda(-u-v+\frac{1}{\kappa^{2}})}{(u+v)^{2}}\right)
^{m+\alpha}\left(-A(2nu+v)+A(u+v)(nu+v)\kappa^{2}-2^{1+n}(u+v)^{2}y^{n}\kappa^{2}
\right.\right.
\end{eqnarray*}
\begin{equation}
\left.\left.\times(-1+(u+v)\kappa^{2})\left(\frac{
\lambda(-u-v+\frac{1}{\kappa^{2}})}{(u+v)^{2}}\right)^{n}\right)\right]
\end{equation}

\vspace{2mm}

\noindent

\begin{eqnarray*}
\partial_{v}{h}=-3\frac{2^{-1-m-n-\alpha}u^{-\alpha}y^{1-m-n}}{\lambda(u+v)^{2}(-1+(u+v)\kappa^{2})^{2}}
\left(\frac{\lambda(-u-v+\frac{1}{\kappa^{2}})}{(u+v)^{2}}\right)^{-m-n-\alpha}
\end{eqnarray*}
\begin{eqnarray*}
\times\left[-2^{n}B(u+v)^{2}y^{n}\kappa^{2}(-1-2m-2\alpha+(u+v)(m+\alpha)\kappa^{2})\left(\frac{\lambda(-u-v+
\frac{1}{\kappa^{2}})}{(u+v)^{2}}\right)^{n}+2^{1+m+\alpha}u^{\alpha}y^{m}\right.
\end{eqnarray*}
\begin{eqnarray*}
\left.\times(-1+(u+v)\kappa^{2})\lambda
\left(\frac{\lambda(-u-v+\frac{1}{\kappa^{2}})}{(u+v)^{2}}\right)
^{m+\alpha}\left(Au(-1+2n-(n-1)(u+v)\kappa^{2})+2^{1+n}(u+v)^{2}y^{n}\kappa^{2}
\right.\right.
\end{eqnarray*}
\begin{equation}
\left.\left.\times(-1+(u+v)\kappa^{2})\left(\frac{
\lambda(-u-v+\frac{1}{\kappa^{2}})}{(u+v)^{2}}\right)^{n}\right)\right]
\end{equation}
\vspace{2mm}

\noindent

\begin{eqnarray*}
\partial_{y}{h}=\frac{1}{\left(u+v\right)\left(-1+\left(u+v\right)\kappa^{2}\right)
\lambda}2^{-1-m-n-\alpha}u^{-\alpha}y^{-m-n}\left(\frac{\left(-u-v+\frac{1}{\kappa^{2}}\right)\lambda}
{\left(u+v\right)^{2}}\right)^{-m-n-\alpha}
\end{eqnarray*}
\begin{eqnarray*}
\left[-3\times
2^{n}B\left(m-1\right)\left(u+v\right)^{2}y^{n}\kappa^{2}\left(\frac{\left(-u-v+\frac{1}{\kappa^{2}}\right)\lambda}
{\left(u+v\right)^{2}}\right)^{n}-2^{1+m+\alpha}u^{\alpha}y^{m}\left(-1+\left(u+v\right)\kappa^{2}\right)\lambda\times\right.
\end{eqnarray*}
\begin{eqnarray*}
\left.\left(\frac{\left(-u-v+\frac{1}{\kappa^{2}}\right)\lambda}
{\left(u+v\right)^{2}}\right)^{m+\alpha}\left\{3A\left(n-1\right)u+2^{n}\left(u+v\right)y^{n}\left(6\left(u+v\right)
\kappa^{2}-7\right)\left(\frac{\left(-u-v+\frac{1}{\kappa^{2}}\right)\lambda}
{\left(u+v\right)^{2}}\right)^{n}\right\}\right]
\end{eqnarray*}

\vspace{2mm}

The Jacobian matrix of the above system is given by,
$$
J_{\left(u,v\right)}^{(RSII)}=\left(\begin{array}{c}\frac{\delta
f}{\delta u} ~~~~~ \frac{\delta f}{\delta v} ~~~~~ \frac{\delta
f}{\delta y}\\ \frac{\delta g}{\delta u}~~~~~ \frac{\delta
g}{\delta v} ~~~~~ \frac{\delta g}{\delta y}\\ \frac{\delta
h}{\delta u} ~~~~~ \frac{\delta h}{\delta v} ~~~~~ \frac{\delta
h}{\delta y}\end{array}\right)
$$

The eigen values of the above matrix are calculated at the
critical point $(u_{c}, v_{c})$ and are found to be~~ ${\bf
\lambda_{1}=5.99118,~~~ \lambda_{2}=-3, ~~~
\lambda_{3}=-2.67056}$. Hence it is a {\bf Saddle point}.

\vspace{2mm}

\subsubsection{\bf NATURE OF COSMOLOGICAL PARAMETERS}

{\bf Deceleration Parameter:}

In this RSII model, the deceleration parameter $q$ can be obtained
as
\begin{equation}\label{rsbasic12}
q^{(RSII)}=-1-\frac{3}{2}\frac{\left\{\frac{\rho}{2\lambda}\left(w_{nvmcg}^{(RSII)}\frac{\rho_{nvmcg}}{\rho}-2\right)-\left(1+w_{nvmcg}^{(RSII)}\frac{\rho_{nvmcg}}{\rho}\right)\right\}}{\left(1+\frac{\rho}{2\lambda}\right)}
\end{equation}
which can be written in terms of dimensionless density parameter
$\Omega_{nvmcg}=\frac{\rho_{nvmcg}}{\rho}$ as in the following
\begin{equation}\label{rsbasic13}
q^{(RSII)}=-1+\frac{3}{2}\frac{\left\{\left(\frac{-\rho}{2\lambda}+1\right)w_{nvmcg}^{(RSII)}\Omega_{nvmcg}+\left(1+\frac{\rho}{\lambda}\right)\right\}}{\left(1+\frac{\rho}{2\lambda}\right)}
\end{equation}
Now since
$\Omega_{nvmcg}=\frac{\rho_{nvmcg}}{\rho}=\frac{u}{u+v}$ and
assuming $\frac{\rho}{\lambda}=\epsilon_{(RSII)}$ we get,
\begin{equation}\label{rsbasic14}
q^{(RSII)}=-1+\frac{3}{2}\frac{\left\{\left(1-\frac{\epsilon_{(RSII)}}{2}\right)w_{nvmcg}^{(RSII)}\frac{u}{u+v}
+\left(1+\epsilon_{(RSII)}\right)\right\}}{\left(1+\frac{\epsilon_{(RSII)}}{2}\right)}
\end{equation}
Considering only the first stable critical point, such that
$(u,v)\rightarrow(u_{1c},v_{1c})$, using (\ref{rsbasic14}) we get,
\begin{equation}\label{rsbasic15}
q_c^{(RSII)}=-1+\frac{3}{2}X_{(RSII)},~~~ where ~~~~
X_{(RSII)}=\frac{\left\{\left(1-\frac{\epsilon_{(RSII)}}{2}\right)w_{nvmcg}^{(RSII)}\frac{u_{1c}}{u_{1c}+v_{1c}}
+\left(1+\epsilon_{(RSII)}\right)\right\}}{\left(1+\frac{\epsilon_{(RSII)}}{2}\right)}
\end{equation}
If
$\epsilon_{(RSII)}=-\frac{2\left[\left(1+w_{nvmcg}^{(RSII)}\right)u_{1c}+v_{1c}\right]}{\left(2-w_{nvmcg}^{(RSII)}\right)u_{1c}+2v_{1c}},
~~X_{(RSII)}=0, ~~~$ we have $q=-1$, which confirms the
accelerated expansion  of the universe. When
$\epsilon_{(RSII)}=-2$ we have $q=-\infty$. Therefore we have
super accelerated expansion of the universe.\\

In this scenario, the Hubble parameter can be obtained as,
\begin{equation}\label{rsbasic16}
H=\frac{2}{3X_{(RSII)}t}
\end{equation}
where the integration constant has been ignored. Integration of
(\ref{rsbasic16}) yields
\begin{equation}\label{rsbasic17}
a(t)=a_0t^{\frac{2}{3X_{(RSII)}}}
\end{equation}
which gives the power law form of expansion of the universe. In
order to have an accelerated expansion of universe in RSII brane
we must have $0<X_{(RSII)}<\frac{2}{3}$. Using this range of
$X_{(RSII)}$ in the equation
$q_{c}^{RSII}=-1+\frac{3}{2}X_{(RSII)}$, we get the range of
$q_{c}^{(RSII)}$ as $-1<q_{c}^{(RSII)}<0$. This is again
consistent with the accelerated expansion of the universe.

{2. \bf Statefinder Parameters }

\vspace{2mm}

\noindent

As so many cosmological models have been developed, so for
discrimination between these contenders, Sahni et al \cite{Sahni1}
proposed a new geometrical diagnostic named the statefinder pair
$\left\{r, s\right\}$. The statefinder parameters are defined as
follows,
\begin{equation}\label{1.1}
r\equiv\frac{\stackrel{...}a}{aH^3},\ \
~~~~~~~~s\equiv\frac{r-1}{3(q-1/2)}
\end{equation}
where $a$ is the scale factor of the universe, $H$ is the Hubble
parameter, a dot denotes differentiation with respect to the
cosmic time $t$.

Here we calculate the statefinder parameters $\{r,s\}$ in order to
get relevant information of DE and DM in the context of background
geometry only without depending on the theory of gravity. The
expressions of the statefinder pair eq.(\ref{1.1}) in the RSII
model can be obtained in the form
\begin{equation}\label{rsbasic19}
r_{(RSII)}=\left(1-\frac{3X_{(RSII)}}{2}\right)\left(1-3X_{(RSII)}\right).
\end{equation}
and
\begin{equation}\label{rsbasic20}
s_{(RSII)}=X_{(RSII)}
\end{equation}

\section{Graphical Analysis}
Graphs are obtained and phase diagrams are drawn in order to
determine the type of critical point obtained in this model. Below
we discuss the results obtained in detail:

The dimensionless density parameters $u$ and $v$ and $y$ are
plotted against each other in figure 1. From the figure we see
that $v$ decreases, and $u$ increases during evolution of the
universe. In figs. 2,3 and 4, the density parameters are plotted
against time. It is evident from the figures that the density of
DM decreases while the density of DE increases as the universe
evolves with an increase in scale factor $a$. So this result is
consistent with the well known idea of an energy dominated
universe. In these figures it is seen that with an increase in the
value of interaction the values of $u$ and $v$ become more and
more comparable to each other, which is quite an expected result.
So we have a possible solution of the cosmic coincidence problem.
A comparative study between \cite{Rudra1,Rudra2,Chowdhury1} and
the current work reveals that DE domination over DM is much less
pronounced in case of NVMCG than in case of other Chaplygin gas
models like, MCG or GCCG. This is a very interesting feature in
the character of NVMCG. Moreover a comparative study of the
current work with \cite{Chowdhury1} reveals the fact that density
of DM is much more comparable to the density of GCCG than to that
of NVMCG. {\bf Hence it can be speculated that NVMCG is perhaps
less effective to play the role of DE in comparison to MCG and
GCCG as well.}

Figs. 5, 6 and 7 shows the phase portrait of the density
parameters of DE and DM. In fig. 5, we see a phase diagram between
density parameters $u(t)$ and $v(t)$. Figs. 6 and 7 shows the
phase diagram between $u(t)$, $y(t)$ and $y(t)$, $v(t)$
respectively. Finally in fig. 8, a 3D-phase diagram between all
the three density parameters is obtained in a single system. As
already stated before that the critical point obtained in this
system is a Saddle point and hence there always remains a question
on the stability of the system. In fig.9, a plot of deceleration
parameter, $q$ is obtained against the EoS parameter, $\omega$. It
is seen that $q$ remains in the negative level thus confirming the
recent cosmic acceleration. Figs. 10 and 11 show the plot of the
statefinder parameters $r$ and $s$ respectively against the EoS
parameter, $\omega$. It is known that in case of $\Lambda$CDM
model $r=1$ and $s=0$. Here we see that in fig. 11 except at
$\omega=0$, the values of $s$ correspond to that of the
$\Lambda$CDM model. In fig. 10, we see that the values of $r$ is
quite different from $1$ corresponding to the values of $\omega$
when $s=0$ in fig. 11.

This gives the deviation of the model from the $\Lambda$CDM model.
Finally in fig. 12, the ratio $v/u$ is plotted against $x=\ln a$.
The decreasing trajectory confirms the existence of an energy
dominated universe with progressive values of scale factor.

\section{Some notes on the mathematical construction of New variable modified Chaplygin gas}
We know that NVMCG is basically an extension of modified Chaplygin
gas, and its mathematical formulation is based on the modification
of the EoS of MCG in a way that makes it more suitable as a
candidate to play the role of DE. Our motive is to investigate how
far successful is NVMCG over MCG as a DE and consequently how far
essential or justified is this modification. So in this section we
consider modified Chaplygin gas as a standard model of dark energy
and perform a comparative study with NVMCG. The EoS for MCG is
given by,
\begin{equation}
p=A\rho_{mcg}-\frac{B}{\rho_{mcg}^{\alpha}}
\end{equation}
where $A$, $B$ and $\alpha$ are positive constants. For negative
pressure we should have
\begin{equation}
\rho_{mcg}<\left(\frac{B}{A}\right)^{\frac{1}{\alpha+1}}
\end{equation}
Now to get negative pressure in case of NVMCG we should have,
\begin{equation}
\rho_{nvmcg}<\left(\frac{B_{0}}{A_{0}}\right)^{\frac{1}{\alpha+1}}a^{\frac{m-n}{\alpha+1}}
\end{equation}

{\bf Case I: for $m=n$}\\
From relation (43), we have,
$\rho_{nvmcg}<\left(\frac{B_{0}}{A_{0}}\right)^{\frac{1}{\alpha+1}}$.
The above value of $\rho_{nvmcg}$ coincides with that of MCG for
$B_{0}=B$ and $A_{0}=A$.

{\bf Case II: for $m>n$}\\
Let $m-n=m_{1}$,where $m_{1}>0$. Therefore relation (43) becomes,
$\rho<\left(\frac{B_{0}}{A_{0}}\right)^{\frac{1}{\alpha+1}}a^{\frac{m_{1}}{\alpha+1}}$.
Now since $a>1$ in an accelerating universe, so
$a^{\frac{m_{1}}{\alpha+1}}>1$. Let
$a^{\frac{m_{1}}{\alpha+1}}=m_{2}$. So we get
$\rho_{nvmcg}<\left(\frac{B_{0}}{A_{0}}\right)^{\frac{1}{\alpha+1}}m_{2}$,
Now since $m_{2}>1$, $\rho_{mcg}<\rho_{nvmcg}$ if $A_{0}=A$ and
$B_{0}=B$. Here it can be seen that the density of NVMCG increases
due to the introduction of scale factor, $a$ in its EoS. It can be
seen from the EoS of NVMCG that this effect increases the
magnitude of the first term and decreases the magnitude of the
second term, which eventually increases the the value of pressure
as a whole. This push towards the towards the positive region in
the value of pressure can be speculated as a basic flaw in the
mathematical construction of NVMCG, since it reduces its
efficiency as a DE. Hence $m>n$ is not at all acceptable as far as
the concept of dark energy is concerned.

{\bf Case III: for $m<n$}\\
In this case $m-n=-m_{1}$. Using this we get,
$\rho<\frac{\left(\frac{B_{0}}{A_{0}}\right)^{\frac{1}{\alpha+1}}}{a^{-\frac{m_{1}}{\alpha+1}}}$.
For an accelerating universe $a>1$, which implies
$a^{\frac{m_{1}}{\alpha+1}}>1$. Therefore
$\rho<\frac{1}{m_{2}}\left(\frac{B_{0}}{A_{0}}\right)^{\frac{1}{\alpha+1}}$.
Now since $m_{2}>1$, which implies $\rho_{mcg}>\rho_{nvmcg}$, if
$A_{0}=A$ and $B_{0}=B$. So there is a relative decrease in the
value of density of NVMCG compared to its counterpart MCG. This in
fact pushes the value of pressure towards the more negative
region, thus enhancing the ability of NVMCG to play the role of
DE. This is the most acceptable case and the EoS of NVMCG should
be constrained by $m<n$.

\section{Study of Future Singularities}
We know that any energy dominated model of the universe will
result in a future singularity. As a result the study of dynamical
model of universe in the presence of DE and DM is in fact
incomplete without the study of these singularities, which are the
ultimate fate of the universe.  It is known that the universe
dominated by phantom energy ends with a future singularity known
as Big Rip \cite{Caldwell1}, due to the violation of dominant
energy condition (DEC). But other than this there are other types
of singularities as well. Nojiri et al \cite{Nojiri1} studied the
various types of singularities that can result from an phantom
energy dominated universe. These possible singularities are
characterized by the growth of energy and curvature at the time of
occurrence of the singularity. It is found that near the
singularity quantum effects becomes very dominant which may
alleviate or even prevent these singularities. So it is extremely
necessary to study these singularities and classify them
accordingly so that we can search for methods to eliminate them.
The appearance of all four types of future singularities in
coupled fluid dark energy, $F(R)$ theory, modified Gauss-Bonnet
gravity and modified $F(R)$ Horava-Lifshitz gravity was
demonstrated in \cite{Nojiri2}. The universal procedure for
resolving such singularities that may lead to bad phenomenological
consequences was proposed. In Rudra et al \cite{Rudra1} it has
been shown that in case of Modified Chaplygin gas(MCG), both Type
I and Type II singularities are possible. However in \cite{Rudra2}
it was shown that GCCG does not result in any type of future
singularity.

\subsection{TYPE I Singularity (Big Rip singularity)}
If $\rho\rightarrow\infty$ , $|p|\rightarrow\infty$ when
$a\rightarrow\infty$ and $t\rightarrow t_{s}$. Then the
singularity formed is said to be the Type I singularity.

In the present case by considering the NVMCG equation of state
from equation (1) we see that when $a\rightarrow\infty$,
$|p|\rightarrow 0$. Therefore we see that there is no possibility
for TypeI singularity, i.e., Big Rip singularity, in case of
NVMCG.

\subsection{TYPE II Singularity (Sudden singularity)}
If $\rho\rightarrow\rho_{s}$ and $\rho_{s}\sim0$, then
$|p|\rightarrow-\infty$ for $t\rightarrow t_{s}$ and $a\rightarrow
a_{s}$, then the resulting singularity is called the Type II
singularity.

Considering the equation of state for NVMCG, We see that if
$\rho_{s}\sim0$, then $|p|\rightarrow -\infty$ for $t\rightarrow
t_{s}$ and $a\rightarrow a_{s}$. Hence there is a strong
possibility of the type II singularity or the sudden singularity
in case of NVMCG.

\subsection{TYPE III Singularity}
For $t\rightarrow t_{s}$, $a\rightarrow a_{s}$,
$\rho\rightarrow\infty$ and $|p|\rightarrow\infty$. Then the
resulting singularity is Type III singularity. It is quite evident
from the equation of state of NVMCG that it supports this type of
singularity.

\subsection{TYPE IV Singularity}
For $t\rightarrow t_{s}$, $a\rightarrow a_{s}$, $\rho\rightarrow0$
and $|p|\rightarrow0$. Then the resulting singularity is Type IV
singularity. Investigation shows that this type of singularity is
not supported by NVMCG type DE.

As a remark, one should stress that our consideration is totally
classical. Nevertheless, it is expected that quantum gravity
effects may play significant role near the singularity. It is
clear that such effects may contribute to the singularity
occurrence or removal too. Unfortunately, due to the absence of a
complete quantum gravity theory only preliminary estimations may
be done.

\noindent

\section{Conclusion}
In this work we have considered New variable modified Chaplygin
gas and tried to determine its efficiency to play the role of dark
energy in an universe described by RSII brane. A numerical system
study was carried out in order to throw some light on the dynamics
of the dark energy. The system was formed and a solution was
obtained. An eigen value analysis of the system at the critical
point showed that the system was far from attaining stability
since it produced a saddle point. Plots were obtained to get a
clear idea about the result of our analysis in a both qualitative
and quantitative aspect. It was found that NNMCG in RSII brane
model is perfectly consistent with the idea of an energy dominated
universe, but the DE domination over DM is much less pronounced in
case of NVMCG when compared to that with Modified Chaplygin gas or
generalized cosmic Chaplygin gas. This is an important result
indeed. It was also discovered from the plots that with the
increase in the magnitude of interaction the values of DE and DM
became more and more comparable to each other thus providing a
solution of the cosmic coincidence problem at higher interaction
scales. The plot of the deceleration parameter revealed that the
model is perfectly consistent with the notion of an energy
dominated universe. The unique trajectories in the plots of
statefinder parameters differentiated the model from other DE
models. An extensive study regarding the mathematical formulation
of NVMCG was performed, and it was found that $m<n$ best suited
the nature of NVMCG as a DE. {\bf Hence we have been able to
constrain the parameters of NVMCG to give the best possible
results}. Finally the future singularities were studied and the
model was found to be affected by some of them, quite unlike the
generalized cosmic Chaplygin gas. In a nutshell it can be said
that the performance of NVMCG as a dark energy is quite moderate.
When compared with other DE models like MCG or GCCG, the weakness
of NVMCG is quite visible. Hence as a conclusion we speculate that
NVMCG is not quite the brightest contender to play the role of DE
and there is a lot of scope for improvement as far as the
mathematical aspect of the model is concerned.

{\bf Acknowledgement:}\\
The authors sincerely acknowledge the facilities provided by the
Inter-University Centre for Astronomy and Astrophysics (IUCAA),
pune, India where a part of the work was carried out.

\end{document}